APOLOGIA PRO VITA SUA: THE VANISHING OF THE WHITE WHALE IN THE MISTS

By

Martin Shubik

July 2018

COWLES FOUNDATION DISCUSSION PAPER NO. 2138

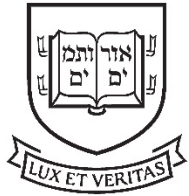

COWLES FOUNDATION FOR RESEARCH IN ECONOMICS
YALE UNIVERSITY
Box 208281
New Haven, Connecticut 06520-8281

http://cowles.yale.edu/

# Apologia pro vita sua:
# The vanishing of the White Whale in the mists
# July 11, 2018
# Martin Shubik

When I finished high school in 1943 it seemed clear to me that the ability to develop and understand the next hundreds of years was going to belong to newer and newer mathematics.  By the time of my graduation from college I knew that unfortunately I had little if any ability to prove formal mathematical theorems, but that I might have some talent in Operations Research and Game Theory thought and the modeling of social phenomena.

There are many analogies among fortune hunting in business, politics and science. The prime task of the gold digger was to go to the Klondikes, find the right mine and mine the richest veins. This task requires motivation, sense of purpose and ability. Techniques and equipment must be developed. Fortune hunting in New England was provided at one time by hunting for whales.  One went to a great whalers' station such as New Bedford and joined the whale hunters. The hunt in academic research is similar. A single minded passion is called for. These notes here are the wrap up comments containing some terminal observations of mine on a hunt for a theory money and financial institutions.

Years ago I decided against writing a formal autobiography, instead I intended to write, on occasion, a more or less casual aide-memoire mixing in a minimal autobiography with a progress report on my own hunt and a sketch of motivation.  I chose as a title for my own work the analogy

"Chasing the White Whale" for my basic pursuit of a theory of money aimed at understanding the vital role of markets, prices and the emergence of money and financial institutions as the carriers of process.

I had wanted to devote my work not only to game theory but to experimental gaming and applications in the behavioral sciences that would appear to be natural topics for this type of treatment. As is indicated in (Shubik, 2016). It took me many years to find the topics that I felt were worth working on, were important and enlightening. In the papers noted (Shubik, 1964, 2016) I have sketched out a rough autobiography and self-justification for what I was trying to do. This note here is written at the end of the trail with a somewhat different viewpoint than I had expected.

The previous papers bring a sketch of my identity up until around 2016. This note covers the small amount that remains.

Possibly one of the extremely difficult reassessments to make is one that recognizes a decay or lowering in one's abilities; but it is a safe bet that after the age of ninety we should proceed with our planning on a daily basis. My close surviving friend, Martin, J. Whitman nearing 93 had played tennis into his nineties provides an example. On December 18, 2017 I had a disastrous fall that damaged my right pelvis causing the loss of a considerable amount of blood, ending a basic ability to walk. A large hematoma took many days to subside. Although I have completely recovered from the specifics of the fall, that single moment signaled a complete change in the nature of my remaining life. Several hours later when I regained consciousness Marty called to give me both condolences and instructions on how to stay the course. I was called the next day with the information that Marty had died from a heart attack.

The combined hell of the hospital together with the purgatory of recovery incarceration enabled me to appreciate that the final exit stage of my life was now. It is not clear this life in this condition is worthwhile, but I try as long as I can still get ½ to 1 page a day without too much pain.

Until my early teens I had no idea of what science or learning was about. I appreciated stories, fiction and history. I had no thoughts about what to do when grown up. Politics in the 1930s and 40s were in the forefront for many who worried about unknown disasters. I knew that many of the "grownups" were worried about the possibility of a major war. In London we knew people from many parts of Europe who were very worried, several from Japan, China, and India, almost all of whom were worried about some form of safety I did not understand. Schools were places to be tolerated as part of "growing up". Movie houses with cartoons helped to separate kids from boring adults.

Until the start of World War II, I had little if any ambitions or understanding of my world beyond school and was hardly aware of differences in societies beyond my family's friends. It took me until going to the University of Toronto and joining the naval reserve to connect with society as a whole. I have recounted elsewhere the only two items of worth that I had been able to glean from life relevant to the future of society and my role. They were the importance of mathematics in the future and its application to the social sciences. By the age of 22 I knew that I had to study and try to develop The Theory of Games.

Although my competences as a debater and political activist were reasonably good I misestimated the social and political importance of both politics and administration in a complex heavily populated world.  I was wrong. I now believe that it is the duty of all individual citizens to devote some portion of their lives to these critical features required to promote and maintain healthy societies and politics.

A combination of choice and circumstances led me to become an academic working in the mathematical social sciences in spite of recognizing that I had little pure mathematical talent. I spent most of my academic life looking for a major problem that I, together with strong mathematicians could tackle. I concluded that the Theory of Money was sufficiently important to understanding many basic aspects of socio-political development. I believe the more than a half of a century I devoted to this endeavor was worthwhile.

It is not for me to judge how well I was able to extend the bounds of our understanding. It remains for me to try to indicate what remains to be done and to try to point out what I believe to be true, but unestablished calling for new work and reassessment.

## The Future

After finishing the technical books on the theory of money and financial institutions in the 2010s it became evident to me that my further work should continue with at least a three-pronged approach. The first was an ongoing assessment of what had been achieved with this new game theoretic approach to money and what remains to be done? This involves noting where and how the socio-political aspects of money enter unavoidably. The topic is essentially institutional. Context matters and opens the abstractions to many natural parameters supplied by the political, legal and social factors of the nations and states. For example, the study of the economics of financing of expenses of epidemic control cannot be specified without the introduction of institutional differences such as the differences in graft levels in the United States, Ghana or China.

The second aspect of future work leads to more philosophical and "iffy" questions. Each of us if sensate and responsible to society as a whole must ask and answer: "If I had my time to do over again would I have acted differently?"

The last and possibly the most difficult task is to assess what of one's previous talents remain and how they can be utilized to point to tasks that help to bring science and society together. My dedicated goal that I pursued almost to the level of insanity was to illustrate and analyze the emergence of markets, prices and money in a complex economy. I called my activity: "The pursuit of the Great White Whale." in the implicit recognition that in the passion and monomania of the pursuit it is dangerous to let the chase itself replace the original goals of the enterprise.

I spent around fifty years dedicated to the study of money and financial institutions and do not regret having done so. I do regret the slowness of my progress. At least I persuaded myself and a few others that my handful of colleagues and I had shown that markets, prices and money emerge as part and parcel of any dynamical political-economy and that economic dynamics of any complex economy is essentially dynamic.

There are several basic questions I had hoped we could answer, but unfortunately I no longer have the talent or time left to settle them to my satisfaction. In particular, given the existence of markets and a money there appear to be only two basic mechanisms for price formation. They are associated with the work of Cournot, Bertrand and Edgeworth. It is my belief that the micro-economic, economic equilibria, and rational expectations rest essentially on quantity or Cournot based strategies and that they provide the capstone for the study of economic statics including Arrow, Debreu, Hicks, Lucas, and others. In contrast the personal price models provide the foundation stone for economic dynamics that underlie the work of Keynes and Schumpeter. The latter are far better suited to deal with reactions to new information.

It is not clear to me that that even by now those who are interested in economic dynamics have a satisfactory definition of what is meant by equilibrium and a proof that it exists in an economy being incessantly bombarded by political, technological, social and other information.

As an admirer of the value of abstract mathematical models in all sciences, I still believe in the virtues of looking for broad general laws that can be applied to many specifics aspects of the world we live in; but in dealing with the political economy it is my current belief that at its best, economics is the hand maiden of socio-political values rather than its ruler.

I chased my whale, and in some ways have had the intellectual pleasures of seeing the powers of trying to construct and analyze models of the political economy; but I have found myself confronting the limits as well as the powers of a mathematical institutional analysis applied to the world we live in.

Even as I felt that I had reached my goal and saw the whale with the harpoon, the mists settled back in. Some insights of the economic powers of markets, money, and prices were gained but the links to the moral sentiments of political economy and society still remain to be understood.

I caught my economic whale. Originally, I thought I would obtain insight into economic welfare and equity, rather than a highly constrained concept of almost mechanistic efficiency of use to human moral governance rather than bureaucratic monetary measures without song or passion. As the vision of the purposeful White Whale appeared, it simultaneously dissolved into a more complex set of moral questions to be answered at best in local time and space with the role of money as a possible accounting aid, but an illusory goal in and of itself.

Money is not the goal nor even a universally good accounting measure for the socio-political economy. The understanding of it together with the array of new financial instruments that have enabled the investigation of strategic information and control has provided new opportunities, to extend the applications of moral philosophy; but without the understanding of and empathy for the purposes at hand offer no more than to provide a more powerful arsenal of economic weaponry to be applied without social purpose.

Further progress requires a joining together of the ad hoc questions from society, history, law and governance with the abstractions of theory. Questions such as "Do matters improve over time?" are probably the wrong questions. Social progress finds equilibrium irrelevant. The question is always what the dynamics are that lead a society from its state today, unto its state tomorrow. The battle between equilibrium and evolution moves on with every answer replaced by more questions. I suspect that the nature of exogenous shock is that the set of economic models is such that convergence to a satisfactory equilibrium fails from many attractive models with  plausible initial conditions. The broad apparently economic optimal equilibrium conditions are local optima where local hill-climbing anti-entropic behavior is confused with global optimization.

RELEVANT REFERENCES

My website in general and. More specifically:

.

Money and Financial Institutions – A Game Theoretic Approach: The Selected Essays of Martin Shubik Volume Two, Cheltenham, UK; Northampton, MA: Edward Elgar Publishing Limited, 1999.

The Theory of Money and Financial Institutions Volume 1, Cambridge, MA; UK: MIT Press, 1999.

The Theory of Money and Financial Institutions Volume 2, Cambridge, MA; UK: MIT Press, 1999.

The Theory of Money and Financial Institutions Volume 3, Cambridge, MA; UK: MIT Press, 2010.

The Guidance of an Enterprise Economy (with D Eric Smith) Cambridge, MA; UK: MIT Press, 2016.[d]